\documentclass[letterpaper, 10 pt, conference]{ieeeconf}

\usepackage{}
\usepackage{mathrsfs}
\usepackage{amsfonts}
\IEEEoverridecommandlockouts
\usepackage{graphicx}
\usepackage{amssymb}
\usepackage{extarrows}
\usepackage{array}
\usepackage{cite}
\usepackage{amsmath}
\usepackage{epstopdf}
\usepackage{enumerate}
\usepackage{xcolor}
\usepackage{soul}
\usepackage{subeqnarray}
\usepackage{cases}
\usepackage{multirow}
\usepackage{breqn}
\usepackage{cite}
\usepackage{caption}
\usepackage{subcaption}
\usepackage[export]{adjustbox}
\usepackage{wrapfig}
\soulregister\cite7
\soulregister\ref7
\newcolumntype{P}[1]{>{\centering\arraybackslash}p{#1}}
\usepackage{color}
\usepackage{caption}
\DeclareMathOperator*{\argmin}{arg\,min}

\makeatletter
\def\mathcolor#1#{\@mathcolor{#1}}
\def\@mathcolor#1#2#3{%
  \protect\leavevmode
  \begingroup
    \color#1{#2}#3%
  \endgroup
}
\makeatother

\begin{document}
\title{Improving Rate of Convergence via Gain Adaptation in Multi-Agent Distributed ADMM Framework }
\author{
        {Towfiq Rahman, Zhihua Qu, and Toru Namerikawa }%
\thanks{T. Rahman and Z. Qu are with Department of Electrical and Computer Engineering, University of Central Florida, USA. T. Namerikawa is with Department of System Design Engineering, Keio
University, Japan. Emails: {\sl towfiq.rahman@knights.ucf.edu}, {\sl qu@ucf.edu} and {\sl namerikawa@sd.keio.ac.jp}. This work was supported in part by the US Department of Energy under awards DE-EE0007998 and DE-EE0009028, and by the US National Science Foundation under grant ECCS-1308928.}%
}
\maketitle

\begin{abstract}
In this paper, the alternating direction method of multipliers (ADMM) is investigated for distributed optimization problems in a networked multi-agent system. In particular, a new adaptive-gain ADMM algorithm is derived in a closed form and under the standard convex property in order to greatly speed up convergence of ADMM-based distributed optimization. Using Lyapunov direct approach, the proposed solution embeds control gains into weighted network matrix among the agents and uses those weights as adaptive penalty gains in the augmented Lagrangian. It is shown that the proposed closed loop gain adaptation scheme significantly improves the convergence time of underlying ADMM optimization. Convergence analysis is provided and simulation results are included to demonstrate the effectiveness of the proposed scheme.

%In this paper, a distributed multi-agent implementation of Alternating Direction Method of Multipliers (ADMM) was studied and an adaptive gain adjusting scheme was proposed for guaranteed faster convergence. The problem was formulated as an interconnected topology of agents having their own individual objective functions to fulfill while satisfying the overall optimization constraints at the network level. Continuous time iterative dynamics for each agent were obtained for the ADMM optimization sub-problems under the assumption of convex properties. Using information transfer between neighboring agents through a network topology, weights on the penalty gain in the augmented Lagrangian can be adaptively modified to further improve the convergence rate of the system. Lyapunov direct method was utilized to provide stability and convergence analysis for the proposed algorithm. Simulation results are included to demonstrate the effectiveness of the gain adaptation scheme to improve the rate of convergence. \\
\end{abstract}


\section{Introduction}
In the era of Internet of Things (IoT) and smart agents, the amount of data available in the network explodes in both size and complexity. In such a multi-agent setting, distributed algorithm for network optimization becomes lucrative and practical since it is not always efficient to gather all the information for a centralized computation\cite{chang2015multi}. The distributed optimization algorithm must be capable of collecting localized data across a connected network of agents, and a common solution must emerge among individual agents without requiring any centralized coordination. In the last decade, the Alternating Direction Method of Multipliers (ADMM) has received much attention due to its ability of decomposing complex optimization problems into a sequence of simpler sub-problems that can be solved asymptotically under certain convex properties \cite{xu2016adaptive}. ADMM was first introduced by Glowinski \& Marroco \cite{glowinski1975approximation} and by Gabay \& Mercier \cite{gabay1976dual}. Most recently, it has been applied to many applications in such areas as image processing \cite{figueiredo2010restoration}, machine learning \cite{forero2010consensus}, resource allocation \cite{joshi2013distributed}, power system optimization \cite{peng2018distributed} etc. These diverse applications also demand a detailed study of ADMM convergence properties \cite{bertsekas1989parallel,boley2013local}.

The convergence speed of ADMM relies on the selection of penalty parameters \cite{ghadimi2015optimal}, which is often manually chosen by the user for a specific problem setup. Convergence rate of ADMM is studied, and earlier work include \cite{bertsekas1989parallel,eckstein1992douglas}. It is now well established in the literature that, if the objective functions are strongly convex and have Lipschitz-continuous gradients, the basic ADMM algorithms have global linear convergence\cite{deng2016global,giselsson2017linear}. The strong convexity conditions are relaxed in \cite{he2015non}, and a constant $O(1/n)$ convergence rate is achieved under mild convex assumptions. It is shown in \cite{kadkhodaie2015accelerated} that convergence can be achieved in $O(1/n^2)$ time if at least one of the objective functions is strongly convex. These specific results all use constant penalty parameters and, in practical applications, efficiency of ADMM is highly sensitive to parameter choices and could be improved via adaptive penalty selection \cite{he2000alternating,nishihara2015general,boyd2011distributed}.

The first approach that comes intuitively is to use different penalty parameters in each iteration. In He et al. \cite{he2000alternating}, an adaptive penalty based on the relative magnitude of primal and dual residuals is proposed to balance their magnitudes. In \cite{ghadimi2015optimal}, primal and dual residuals are also used to improve a defined convergence factor while solving a class of quadratic optimization problem using ADMM. In both these cases, the ADMM algorithm is shown to converge, but global computation of primal and dual residuals are required and hence the resulting algorithm is no longer distributed. In \cite{makhdoumi2017convergence}, distributed ADMM is implemented to minimize locally known convex functions over a network, and the effect of communication weights on the convergence rate is investigated. In \cite{qu2014cooperative}, the weighted network matrix is adaptively tuned to improve convergence in a consensus-based distributed problem framework using cooperative control. This idea is used in \cite{song2016fast} where a consensus based distributed ADMM is formulated with a predefined network structure, for which primal and dual residuals are balanced locally by each agent. However, their adaptive penalty needs to be reset after several iterations to guarantee convergence, which results in much weakened convergence conditions.  More recently, adaptive penalty parameters are used \cite{xu2016adaptive} to improve convergence speed by estimating the local curvatures of the dual functions. However, as pointed out in \cite{xu2017adaptive}, an increase in the number of nodes causes the local curvature estimation to be inaccurate and possibly unstable. 

In this paper, a Lyapunov-based analytical design methodology is proposed to synthesize adaptive penalty parameters for ADMM to ensure convergence and improve convergence time, all in a multi-agent setting. The proposed distributed ADMM algorithm is designed in four steps. First, distributed control gains are embedded into a row-stochastic weighted network connectivity matrix to ensure consistency of the ADMM between its constraints and network connectivity. Second, the entries of weighted network matrix are embedded as the penalty parameters into the augmented Lagrangian for ADMM so they can be adjusted in a distributed  manner for each agent to use its local information and optimize its local objective function. Third, utilizing the convex property of the individual agents' objective functions, the standard ADMM formulation is applied to the newly formulated augmented Lagrangian, the resulting ADMM algorithm with adaptive gains is shown to be asymptotically convergent, and its iterative ADMM updating laws are derived. Fourth, using the Lyapunov direct method, adaptive gain updating laws are analytically synthesized, and the improvement of convergence is proven.

The rest of the paper is organized as follows. In section II, %the distributed optimization problem is formulated, and 
the proposed adaptive-gain ADMM algorithm is developed. In section III, the main results of gain adaptation and convergence improvement are established. Superior performance of the proposed ADMM algorithm is shown through comparative simulation studies in section IV. And, section V contains the conclusions.

\section{Problem Formulation}
In this section, a general multi-agent distributed optimization problem is presented for a network of agents and then converted into the ADMM framework. Based on network connectivity and the convex property of objective functions, optimum search iterates at each agent are derived to solve the ADMM sub-problems. Using these iterates and local communication among agents, an adaptive gain updating algorithm is synthesized to further improve convergence of the ADMM algorithm to an optimal solution.

\subsection{ADMM Based Distributed Algorithm}
Let us consider the following distributed optimization problem with a conforming communication topology among agents:
\begin{subequations}
\begin{eqnarray}
\min && \sum_{i \in \mathcal{N}}f_i(x_i) \label{value} \\
\text{s.t.}  && \sum_{j\in \mathcal{N}_i}  {A_{ij}} x_{j} = 0  \hspace{1.5em} \text{for}\hspace{1.5em} i \in \mathcal{N},
\label{linearConstraint}
\end{eqnarray}
\label{DistOpt}
\end{subequations}

\vspace*{-0.1in}

\noindent where $\mathcal{N} = \{1,2,\cdots, N\}$ is the set of agents. For agent $i$, $\mathcal{N}_i$ denotes the set of its neighbors including itself, $x_i \in \mathbb{R}^n$ is its state vector, $f_i(x_i)$ is its objective function, and $A_{ij}$ are matrices of appropriate dimensions in the linear constrained equations representing the interconnection of the physical layer. The problem can be perceived as each agent $i$ optimizing its own objective function $f_i(x_i)$ while satisfying the interconnection constraint of \eqref{linearConstraint}. The following assumption is made on the individual objective functions.

\indent \textbf{Assumption 1}: \textit{Functions $f_i$, $i\in \mathcal{N}$, are convex and differentiable, and their gradients denoted by $\nabla_{x_i}f_i(x_i)$ are locally Lipschitz. The set of optimal solutions to \eqref{DistOpt} is not empty, and the corresponding minimum of \eqref{value} is finite.}

\subsection{Network of Agents}
\begin{figure}
\begin{center}
\includegraphics[width = 6cm, height = 4cm]{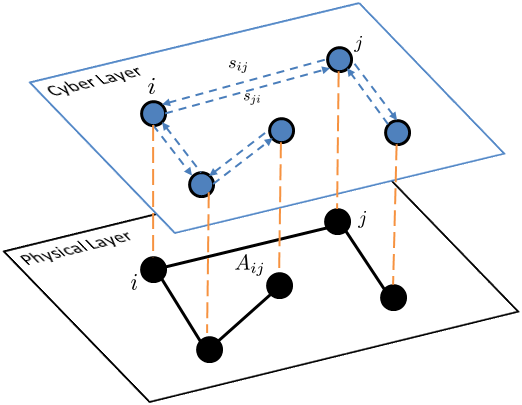}
\caption{\small{Networked cyber-physical system of multi agents.}}\label{fig:CPS}
\end{center}
\vspace{-1em}
\end{figure}

Local communication in the network is characterized by a bidirectional graph $\mathcal{G} = (\mathcal{N},\mathcal{E})$; specifically, its sensing/communication matrix is binary and of form \cite{qu2009cooperative}:
\begin{equation}
S = \left[\begin{array}{cccc}
1  & s_{12} & \cdots & s_{1 N}\\
\vdots  & \vdots & \vdots &	\vdots \\
s_{N1}& s_{N 2} & \cdots & 1
\end{array}\right].
\label{CommunicationMatrix}
\end{equation}
Matrix $S$ has 1 in the diagonal as every agent knows its own information, and it is equal to the sum of the adjacency matrix and identity matrix. The following assumption ensures conformity and connectivity of the network, and the multi-agent system is visualized in figure \ref{fig:CPS}.
\\
\indent \textbf{Assumption 2}: \textit{The communication graph conforms with system constraints in the sense that, if $A_{ij} \neq 0$ or $A_{ji} \neq 0$, $s_{ij} = s_{ji} = 1$. And, the communication graph is connected, i.e., matrix $S$ is irreducible.}

\subsection{Adaptive-Gain ADMM}

To solve \eqref{DistOpt} using ADMM, we introduce a set of auxiliary variables, $z_{ji}$, which are the estimates of agent $j$'s variables by agent $i$ \cite{peng2018distributed}. Then, problem \eqref{DistOpt} can be restated as
%\vspace{-1em}
\begin{subequations}
\begin{eqnarray}
\min && \sum_{i \in \mathcal{N}}f_i(x_i)\\
\text{s.t.}  &&\sum_{j \in \mathcal{N}_i}
{A_{ij}}
z_{ji} = 0 \hspace{1.5em}  \text{for}\hspace{1.5em} i \in \mathcal{N} \label{ADMMConstraint}\\
&&x_i =  z_{ij} \qquad j \in \mathcal{N}_i, \;\; i \in \mathcal{N} \label{eq:consensus}
\end{eqnarray}
\label{ADMMFormulation}
\end{subequations}
\vspace{-2em}

\noindent where $z_{ij}$ are relaxation variables used in the standard ADMM \cite{boyd2011distributed}.  All the agents related to the $i$th agent try to make their estimates $z_{ij}$ of  $x_i$ reach consensus so that a solution to \eqref{ADMMFormulation}
converges to an optimal solution $x^*$ of the original problem \eqref{DistOpt}.
The goal of this reformulation is to solve the optimization problem in a distributed fashion that agent $i$ solves its own optimization sub-problem by exchanging information with its neighboring nodes in set $\mathcal{N}_i$. To this end, we form the so-called augmented Lagrangian as:
\begin{equation}
 L_{D}(x,z,\lambda,\mu) = \sum_{i \in \mathcal{N}}L_i(x_i,z_{ij},\lambda_{ij},\mu_{i}), \label{eq:Lagrangian}
\end{equation}
\noindent where $\lambda_{ij}$ and $\mu_i\; (j \in \mathcal{N}_i,i \in \mathcal{N}_i)$, are the Lagrange multipliers (dual variables) associated with the constraints,
\begin{align}
L_i =& f_i(x_i) +  \sum_{j \in \mathcal{N}_i} \Big[\lambda^T_{ij}(x_i - z_{ij}) +  \frac{d_{ij}}{2}||x_i-z_{ij}||^2\Big] \nonumber \\
& + \mu_{i}^T\sum_{j \in \mathcal{N}_i}A_{ij}z_{ji} + \frac{w_{i}}{2}\Big|\Big|\sum_{j \in \mathcal{N}_i}A_{ij}z_{ji}\Big|\Big|^2,  \nonumber
\end{align}
and $d_{ij} \geq 0$ are regularized but time-varying penalty parameters from a row-stochastic gain matrix $D^k$, and $w_i>0$ is the penalty parameter associated with constraint \eqref{ADMMConstraint}. The augmented Lagrangian reduces to a standard Lagrangian $L_0$ when the penalty terms are removed (i.e., $d_{ij} =0$ for all $i \in \mathcal{N}$ and $j \in \mathcal{N}_i$).

To ensure that the proposed adaptive scheme is consistent with local communication network, gain matrix $D^k\in \mathbb{R}^{n \times n}$ are locally calculated (by row) as
\begin{equation}
D^k = \left[d_{ij}^k\right], \;\;\;
d_{ij}^k= \frac{s_{ij}\beta_{ij}^k}{\sum_{l=1}^n s_{il}\beta_{il}^k},
\label{fixedpenalty}
\end{equation}
where $k \in \aleph^+$ is the discrete time step, and  $\beta_{ij}^k\geq0$ as local scalar gains (with initial gain values of $\beta_{ij}^0>0$). Entries $d_{ij}^k$ (or equivalently $\beta_{ij}^k$) will be updated real-time according to the proposed design. Clearly, gain matrix $D^k$ is non-negative,  row-stochastic and diagonally positive. The proposed adaptive ADMM approach naturally lends itself to distributed optimization and gives us the flexibility of adjusting the gains on received information. Should all $d_{ij}^k$ become a constant penalty parameter $\rho$, the proposed design reduces to the standard ADMM algorithm \cite{boyd2011distributed}.

The ADMM algorithm consists of an $x$-minimization step, a $z$-minimization step, and an update of dual variables. The proposed ADMM algorithm is obtained by applying these steps to the above reformulation, that is,
\begin{subequations}
\begin{enumerate}
    \item For any $i \in \mathcal{N}$, $x_i$ is updated according to
    \begin{equation}
        x_i^{k+1} := \argmin_{x_i \in \mathbb{R}^n} L_D(x^k,z^k,\mu^k,\lambda^k)
        \label{x-problem}
    \end{equation}
    \item For any $i \in \mathcal{N}$ and for $j \in \mathcal{N}_i$, $z_{ji}$ is solved as
    \begin{align}
        z_{ij}^{k+1} :=&  \argmin_{z_{ij}\in \mathbb{R}^n} \hspace{0.5em} L_D(x^{k+1},z^k,\mu^k,\lambda^k)
        \label{z-problem}
    \end{align}
    \item For any $i \in \mathcal{N}$ and for $j \in \mathcal{N}_i$, $\mu_i$ and $\lambda_{ij}$ evolves as
    \begin{eqnarray}
        \mu_i^{k+1} := & \mu_{i}^k + w_i\sum_{j \in \mathcal{N}_i} A_{ij}z_{ji}^{k+1}\label{mu-problem}\\
        \lambda_{ij}^{k+1} := & \lambda_{ij}^k +  d_{ij}^k\Big[x_i^{k+1}- z_{ij}^{k+1}\Big]\label{lambda-problem}
    \end{eqnarray}
\end{enumerate}
\label{ADMM-problem}
\end{subequations}
Convergence property of the proposed  ADMM algorithm \eqref{ADMM-problem} for primal-dual sequences of $\{x_i^k,z_{ij}^k\}$ and $\{\lambda_{ij}^k,\mu_{i}^k\}$
is summarized as the following lemma, and its proof included in Appendix extends the existing ADMM results to version \eqref{ADMM-problem} with time-varying gains.

\textbf{Lemma 1}: \textit{Under assumptions 1 and 2, ADMM algorithm \eqref{ADMM-problem} is convergent to an optimal solution.}

\subsection{Iterative Laws of ADMM}

Under assumption 1, the $i$th agent can solve the sub-optimization problems in \eqref{ADMM-problem} iteratively. In particular, the gradient descent technique can be applied to the $x$-minimization step, while the $z$-minimization step has a closed form solution once $x_i^{k+1}$ is determined. Hence, the ADDM iterative algorithm becomes: for agent $i$,
\begin{subequations}
\begin{align}
&\hat{x}_i^{k+1} =  \hat{x}_i^k -\alpha_i \Big[\nabla_{\hat{x}_i} f_i(\hat{x}_i^k)+ \sum_{j \in \mathcal{N}_i} \big[ \hat{\lambda}_{ij}^k + d_{ij}^k(\hat{x}_i^k- \hat{z}_{ij}^k)\big]\Big]
\label{x-update}
%\nonumber
\\
& \hat{z}_{ji}^{k+1} =  \hat{x}_j^{k+1} + \frac{1}{d_{ji}^k}\Big[\hat{\lambda}_{ji}^k - A_{ij}^T\hat{\mu}_i^k - w_i^kA_{ij}^T\sum_{\phi \in \mathcal{N}_i}A_{i\phi}\hat{z}_{\phi i}^k\Big]
\label{z-update}
%\nonumber
\\
&\hat{\mu}_{i}^{k+1} =  \hat{\mu}_{i}^k +  w_i\sum_{j \in \mathcal{N}_i} A_{ij}\hat{z}_{ji}^{k+1} \hspace{3em}
\label{mu-update}
%\nonumber
\\
&\hat{\lambda}_{ji}^{k+1} = \hat{\lambda}_{ji}^k +  d_{ji}^k(\hat{x}_j^{k+1}- \hat{z}_{ji}^{k+1})
\label{lambda-update}
\end{align}
\label{dynamic}
\end{subequations}
\noindent where $0 < \alpha_i << 1$ is the step size. The iterative algorithm \eqref{dynamic} is a gradient-based optimization and, under convexity, it is guaranteed that its trajectory moves closer to an optimal solution. Hence, the convergence proof of algorithm \eqref{dynamic} is neglected for briefness.

\begin{figure}
\begin{center}
\includegraphics[width = 7cm, height = 4cm]{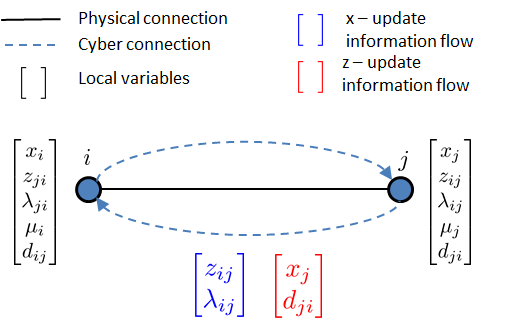}
\caption{\small{Agent $i$'s information flow for local variable update.}}\label{fig:update}
\end{center}
\vspace{-1em}
\end{figure}

In the implementation, agent $i$ updates not only its own state vector $x_i$ but also estimates $z_{ji}$ of its neighboring agents' states as well as the associated Lagrange multipliers $\lambda_{ji}$ and $\mu_i$. This information flow for updating the iterates are shown in figure \ref{fig:update}.

In most of the existing ADMM literature, penalties $d_{ij}^k$ are set to be constant and identical\cite{deng2016global,giselsson2017linear,he2015non,kadkhodaie2015accelerated}. Advantage of using adaptive penalty is noted in \cite{he2000alternating,nishihara2015general,boyd2011distributed},  but those results either require global information or have convergence and scalability issues. These motivate us to develop the proposed adaptive penalty algorithm whose gain matrix is chosen constructively to retain the ADMM's distributed nature while enhancing its scalability and convergence. Specifically,  the $i$th agent dynamically adjusts its penalties (i.e., the $i$th row entries of matrix $D^k$) to improve convergence time of ADMM. This design objective is achieved through making the value of an appropriate Lyapunov function decrease more at each of the iteration steps. This idea was first applied successfully to cooperative control among a network of cooperative agents in \cite{qu2014cooperative}. An application of this idea to ADMM is pursued in the next section.

\section{Improvement of convergence rate via adaptive gain}
At each of the iteration steps, convergence of ADMM algorithm \eqref{dynamic} can be measured using the following Lyapunov function by agent $i$:
\begin{align}
     E_i^k =& ||\tilde{x}_i^k||^2  + ||\tilde{\mu}_i^k||^2
      + \sum_{j \in \mathcal{N}_i} \Big[||\tilde{z}_{ji}^k||^2 + ||\tilde{\lambda}_{ji}^k||^2 \Big],
      \label{soloLyapunov}
\end{align}
where $\tilde{x}^{k} = \hat{x}^{k} - \hat{x}^{k-1}$, $\tilde{z}^k = \hat{z}^k - \hat{z}^{k-1}$, $\tilde{\lambda}^k = \hat{\lambda}^k - \hat{\lambda}^{k-1}$ and $\tilde{\mu}^k = \hat{\mu}^k - \hat{\mu}^{k-1}$ are incremental residues of the primal and dual variables. The following theorem provides the proposed distributed adaptive-gain algorithm, and its proof is included in the Appendix.

\textbf{Theorem 1:} \textit{Convergence of ADMM algorithm \eqref{dynamic} is improved if $E_i^{k+1}$ is made to be less through locally and adaptively choosing $\beta_{ij}^{k}$. Specifically, for each of $k \in \aleph^+$, only two of the penalties $d_{ij}^{k}$ (equivalently, gains $\beta_{ij}^k$) are adaptively adjusted as}
\begin{equation}
d_{il_i}^{k} = d_{il_i}^{k-1} + \epsilon_i^k \qquad d_{im_i}^{k} = d_{im_i}^{k-1} - \epsilon_i^k, \label{ADMM-gains}
\end{equation}
\textit{where indices $l_i$ and $m_i$ are determined according to}
\begin{align*}
&l_i \in \mathcal{N}_i \Longrightarrow [\hat{x}_i-\hat{z}_{il_i}] = \max_{j \in \mathcal{N}_i}[(\nabla_{\hat{x}_i}f_i(\hat{x}_i^k)\bullet (\hat{x}_i^k-\hat{z}_{ij}^k)], \\
&m_i \in \mathcal{N}_i \Longrightarrow [\hat{x}_i-\hat{z}_{im_i}] = \min_{j \in \mathcal{N}_i}[(\nabla_{\hat{x}_i}f_i(\hat{x}_i^k)\bullet (\hat{x}_i^k-\hat{z}_{ij}^k)],
\end{align*}
quantity
\begin{equation} h_i(\tilde{x},\tilde{z}) = 2\alpha_i\tilde{x}_i^k\Big[(\tilde{x}_i^k - \tilde{z}_{il_i}^k) - (\tilde{x}_i^k - \tilde{z}_{im_i}^k) \Big] \label{ADMM-h}
\end{equation}
\textit{is calculated using the locally-available information, and adjustment $\epsilon_i$ is chosen to be: for some $0<\gamma_i<1$,}
\begin{equation}
\epsilon_i^k = \left \{
\begin{array} {llll}
\vspace{1em}
\gamma_id_{im_i}^{k-1} &\text{if} \; h_i(\tilde{x},\tilde{z}) >0 \;  \\
\vspace{1em}
-\gamma_id_{il_i}^{k-1}& \text{if} \; h_i(\tilde{x},\tilde{z}) < 0 \; \\
0& \text{otherwise.}
\end{array}\right. \label{ADMM-epsilon}
\end{equation}
%It should be noted that all the information required for the $i$th agent to locally calculate and evaluate $h_i(x,z)$ are already gathered from its neighbors. From lemma 1 and theorem 1, we know that the proposed adaptation always improves convergence. The figures obtained in the simulation will further explain and help visualize the concept presented.

\section{Simulation results}
In this section, the proposed gain adaptation technique is illustrated through simulations and in two parts. First, the time trajectory of convergence error measure under the proposed adaptive-gain ADMM is compared to that under fixed penalties for a $5$-agent network. Second, comparative studies are done for scaled-up networks up to $100$ agents and for different network topologies so improvements of convergence speed are established together with scalability. In both cases, the error residual in the form of $\max_i\sum_{j \in \mathcal{N}_i}|x_i-z_{ij}|$ is chosen to measure convergence, and the tolerance threshold of $1\times 10^{-4}$ is used to either stop the simulation or determine the number of iterations required for convergence when comparing the ADMM algorithms. In the implementation of algorithms, the following choices are made: $\alpha_i=0.1$, $w_i=1$, $f_i(x_i)$ is convex and only known to the $i$th agent. As the base case, a ring-like topology where each agent is connected to two other agents is constructed. Then, two other topologies are generated on top of the ring-structure where agents are randomly interlinked up to a maximum of five other agents. For each of the scenarios generated, five sets of simulations with different initial conditions are run.

\begin{figure}[!htbp]
\begin{center}
\includegraphics[width = 9cm, height = 4cm]{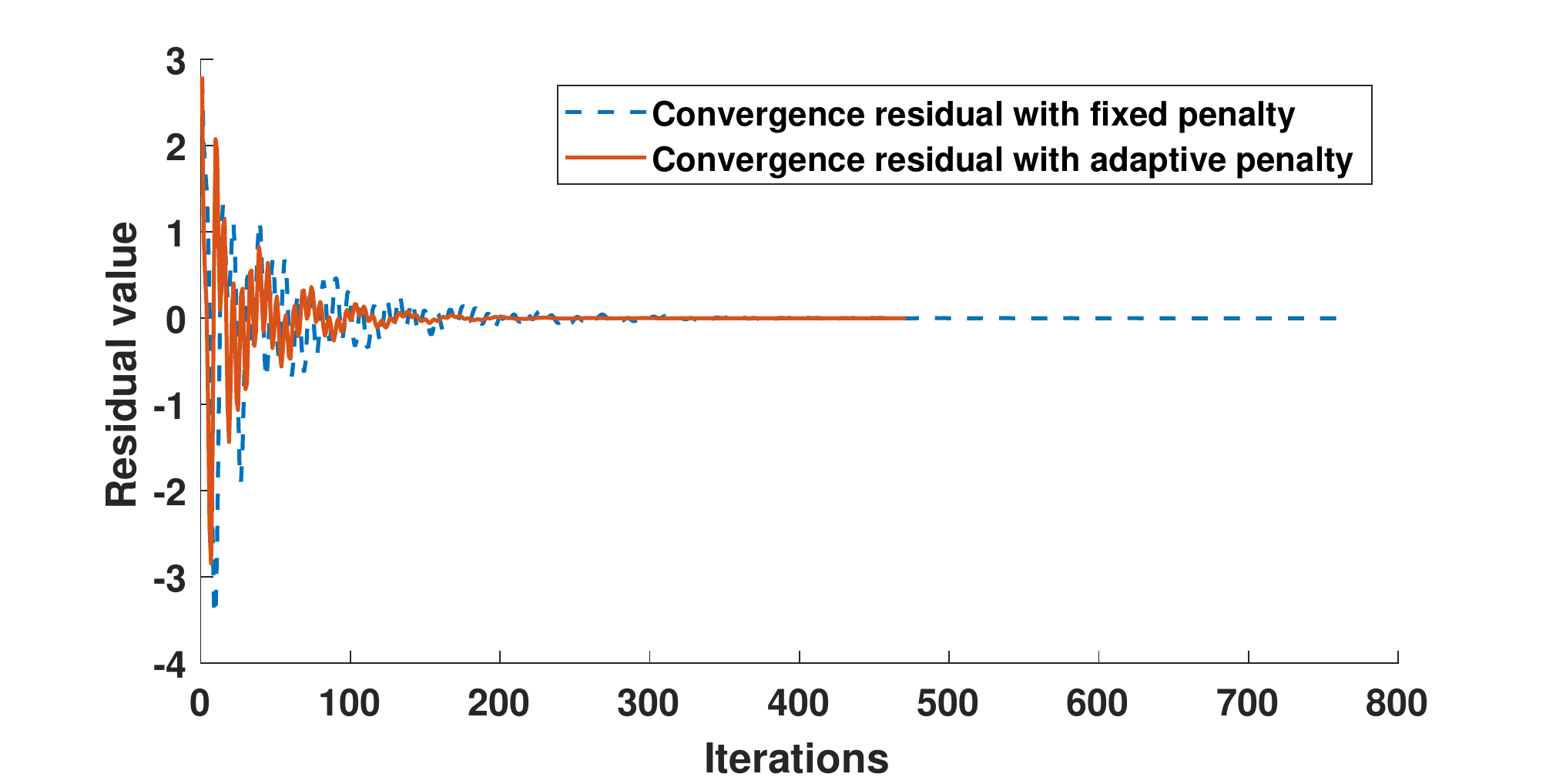}
\caption{\small{Convergence errors: fixed again versus adaptive penalty}}\label{fig:fixedAdaptive}
\end{center}
\end{figure}

First, lets begin with the $5$-agent ring network (whose connectivity matrix is cyclic). The iterative ADMM algorithm \eqref{dynamic} is implemented for each agent, and simulations are run twice: one with fixed gains (in which case $d_{ij}$ is computed using \eqref{fixedpenalty} at $k=0$ and then kept constant), and another with adaptive gains (whose initial values are calculated the same way and then the gains are updated over time according to theorem 1). Comparison of convergence under adaptive-gain ADMM versus fixed-penalty ADMM is shown in figure \ref{fig:fixedAdaptive}.

In the second set of simulation studies, the network and its scaled-up versions up to $100$ agents are simulated for various initial conditions and with different connected network topologies as described above. Again, each simulation setup is repeated twice: one of constant penalty, and another with adaptive gains. For a network of certain agents, convergence times are recorded for different initial conditions and randomly generated network topologies, and their average is recorded. The iteration limit is set to $30,000$. Table \ref{table:comparative} provides comparative summary results for networks of different sizes.

\begin{table}[!htbp]
\begin{center}
\caption{Comparative analysis of the algorithms}\label{tab1}
%\begin{tabular}{|p{2.2cm}||p{1.5cm}|p{1.5cm}|}
\begin{tabular}{|c||c|c|}
\hline
\multicolumn{3}{|c|}{Iterations required for convergence}  \\
\hline
 Number of agents & Fixed penalty & Adaptive penalty \\
 \hline
 $5$ & 765 & 471 \\
 \hline
 $10$ & 1138 & 573 \\
 \hline
 $25$ & 5934 & 1924 \\
 \hline
 $50$ & 11968 & 7342 \\
 \hline
 $100$ & Exceeds iteration limit & 20369 \\
 \hline
\end{tabular}
\label{table:comparative}
\end{center}
\end{table}

The results of our two-part studies clearly show convergence improvements for the proposed adaptive ADMM algorithm.

\section{Conclusion}
In this paper, a distributed multi-agent ADMM algorithm with adaptive gains is developed. The convex properties are utilized to obtain closed form iterative dynamics for the optimization sub-problems. In contrast to the standard ADMM which uses a fixed penalty gain in the augmented Lagrangian, the proposed algorithm embeds control gains into a row-stochastic matrix based on network connectivity, utilizes the matrix coefficients as the penalty parameters in ADMM, and uses information received by each agent from its neighbors to adaptively adjust these penalties. The proposed adaptive algorithm is both distributed and of closed form, and it substantially improves the rate of ADMM agents' convergence to an optimal solution. The improvement is analytically shown by the Lyapunov direct approach. Numerical simulation demonstrates the effectiveness of the proposed adaptive-gain ADMM.

\bibliographystyle{IEEEtran}
\bibliography{IEEEabrv,AdaptiveGainConvergence}

\begin{thebibliography}{10}
\providecommand{\url}[1]{#1}
\csname url@samestyle\endcsname
\providecommand{\newblock}{\relax}
\providecommand{\bibinfo}[2]{#2}
\providecommand{\BIBentrySTDinterwordspacing}{\spaceskip=0pt\relax}
\providecommand{\BIBentryALTinterwordstretchfactor}{4}
\providecommand{\BIBentryALTinterwordspacing}{\spaceskip=\fontdimen2\font plus
\BIBentryALTinterwordstretchfactor\fontdimen3\font minus
  \fontdimen4\font\relax}
\providecommand{\BIBforeignlanguage}[2]{{%
\expandafter\ifx\csname l@#1\endcsname\relax
\typeout{** WARNING: IEEEtran.bst: No hyphenation pattern has been}%
\typeout{** loaded for the language `#1'. Using the pattern for}%
\typeout{** the default language instead.}%
\else
\language=\csname l@#1\endcsname
\fi
#2}}
\providecommand{\BIBdecl}{\relax}
\BIBdecl

\bibitem{chang2015multi}
T.-H. Chang, M.~Hong, and X.~Wang, ``Multi-agent distributed optimization via
  inexact consensus {ADMM},'' \emph{IEEE Transactions on Signal Processing},
  vol.~63, pp. 482--497, 2015.

\bibitem{xu2016adaptive}
Z.~Xu, M.~A. Figueiredo, and T.~Goldstein, ``Adaptive {ADMM} with spectral
  penalty parameter selection,'' \emph{arXiv:1605.07246}, 2016.

\bibitem{glowinski1975approximation}
R.~Glowinski and A.~Marroco, ``Sur l'approximation, par {\'e}l{\'e}ments finis
  d'ordre un, et la r{\'e}solution, par p{\'e}nalisation-dualit{\'e} d'une
  classe de probl{\`e}mes de dirichlet non lin{\'e}aires,'' \emph{Revue
  Fran{\c{c}}aise d'Automatique, Informatique, Recherche Op{\'e}rationnelle.
  Analyse Num{\'e}rique}, vol.~9, pp. 41--76, 1975.

\bibitem{gabay1976dual}
D.~Gabay and B.~Mercier, ``A dual algorithm for the solution of nonlinear
  variational problems via finite element approximation,'' \emph{Computers \&
  Mathematics with Applications}, vol.~2, pp. 17--40, 1976.

\bibitem{figueiredo2010restoration}
M.~A. Figueiredo and J.~M. Bioucas-Dias, ``Restoration of poissonian images
  using alternating direction optimization,'' \emph{IEEE Transactions on Image
  Processing}, vol.~19, pp. 3133--3145, 2010.

\bibitem{forero2010consensus}
P.~A. Forero, A.~Cano, and G.~B. Giannakis, ``Consensus-based distributed
  support vector machines,'' \emph{Journal of Machine Learning Research},
  vol.~11, pp. 1663--1707, 2010.

\bibitem{joshi2013distributed}
S.~Joshi, M.~Codreanu, and M.~Latva-aho, ``Distributed {SINR} balancing for
  {MISO} downlink systems via the {A}lternating {D}irection {M}ethod of
  {M}ultipliers,'' in \emph{Modeling \& Optimization in Mobile, Ad Hoc \&
  Wireless Networks}, 2013, pp. 318--325.

\bibitem{peng2018distributed}
Q.~Peng and S.~H. Low, ``Distributed optimal power flow algorithm for radial
  networks 1: Balanced single phase case,'' \emph{IEEE Transactions on Smart
  Grid}, vol.~9, pp. 111--121, 2018.

\bibitem{bertsekas1989parallel}
D.~P. Bertsekas and J.~N. Tsitsiklis, \emph{Parallel and Distributed
  Computation: Numerical Methods}.\hskip 1em plus 0.5em minus 0.4em\relax
  Prentice Hall Englewood Cliffs, NJ, 1989.

\bibitem{boley2013local}
D.~Boley, ``Local linear convergence of the {A}lternating {D}irection {M}ethod
  of {M}ultipliers on quadratic or linear programs,'' \emph{SIAM Journal on
  Optimization}, vol.~23, pp. 2183--2207, 2013.

\bibitem{ghadimi2015optimal}
E.~Ghadimi, A.~Teixeira, I.~Shames, and M.~Johansson, ``Optimal parameter
  selection for the alternating direction method of multipliers ({ADMM}):
  quadratic problems,'' \emph{IEEE Transactions on Automatic Control}, vol.~60,
  pp. 644--658, 2015.

\bibitem{eckstein1992douglas}
J.~Eckstein and D.~P. Bertsekas, ``On the {D}ouglas-{R}achford splitting method
  and the proximal point algorithm for maximal monotone operators,''
  \emph{Mathematical Programming}, vol.~55, pp. 293--318, 1992.

\bibitem{deng2016global}
W.~Deng and W.~Yin, ``On the global and linear convergence of the generalized
  {A}lternating {D}irection {M}ethod of {M}ultipliers,'' \emph{Journal of
  Scientific Computing}, vol.~66, pp. 889--916, 2016.

\bibitem{giselsson2017linear}
P.~Giselsson and S.~Boyd, ``Linear convergence and metric selection for
  {D}ouglas-{R}achford splitting and {ADMM},'' \emph{IEEE Transactions on
  Automatic Control}, vol.~62, pp. 532--544, 2017.

\bibitem{he2015non}
B.~He and X.~Yuan, ``On non-ergodic convergence rate of {D}ouglas-{R}achford
  alternating direction method of multipliers,'' \emph{Numerische Mathematik},
  vol. 130, pp. 567--577, 2015.

\bibitem{kadkhodaie2015accelerated}
M.~Kadkhodaie, K.~Christakopoulou, M.~Sanjabi, and A.~Banerjee, ``Accelerated
  alternating direction method of multipliers,'' in \emph{Proceedings of the
  21st ACM SIGKDD International Conference on Knowledge Discovery and Data
  Mining}, 2015, pp. 497--506.

\bibitem{he2000alternating}
B.~He, H.~Yang, and S.~Wang, ``Alternating direction method with self-adaptive
  penalty parameters for monotone variational inequalities,'' \emph{Journal of
  Optimization Theory and Applications}, vol. 106, pp. 337--356, 2000.

\bibitem{nishihara2015general}
R.~Nishihara, L.~Lessard, B.~Recht, A.~Packard, and M.~I. Jordan, ``A general
  analysis of the convergence of {ADMM},'' \emph{arXiv preprint
  arXiv:1502.02009}, 2015.

\bibitem{boyd2011distributed}
S.~Boyd, N.~Parikh, E.~Chu, B.~Peleato, J.~Eckstein \emph{et~al.},
  ``Distributed optimization and statistical learning via the {A}lternating
  {D}irection {M}ethod of {M}ultipliers,'' \emph{Foundations and Trends in
  Machine learning}, vol.~3, pp. 1--122, 2011.

\bibitem{makhdoumi2017convergence}
A.~Makhdoumi and A.~Ozdaglar, ``Convergence rate of distributed {ADMM} over
  networks,'' \emph{IEEE Transactions on Automatic Control}, vol.~62, no.~10,
  pp. 5082--5095, 2017.

\bibitem{qu2014cooperative}
Z.~Qu, C.~Li, and F.~Lewis, ``Cooperative control with distributed gain
  adaptation and connectivity estimation for directed networks,''
  \emph{International Journal of Robust and Nonlinear Control}, vol.~24, pp.
  450--476, 2014.

\bibitem{song2016fast}
C.~Song, S.~Yoon, and V.~Pavlovic, ``Fast {ADMM} algorithm for distributed
  optimization with adaptive penalty.'' in \emph{Association for the
  Advancement of Artificial Intelligence}, 2016, pp. 753--759.

\bibitem{xu2017adaptive}
Z.~Xu, G.~Taylor, H.~Li, M.~Figueiredo, X.~Yuan, and T.~Goldstein, ``Adaptive
  consensus {ADMM} for distributed optimization,'' \emph{Proceedings of the
  34th International Conference on Machine Learning, vol.70, JMLR.org}, 2017.

\bibitem{qu2009cooperative}
Z.~Qu, \emph{Cooperative Control of Dynamical Systems}.\hskip 1em plus 0.5em
  minus 0.4em\relax Springer, 2009.

\bibitem{boyd2004convex}
S.~Boyd and L.~Vandenberghe, \emph{Convex Optimization}.\hskip 1em plus 0.5em
  minus 0.4em\relax Cambridge University Press, 2004.

\end{thebibliography}

\section*{Appendix}
\vspace*{-0.1in}
\subsection{Proof of Lemma 1}
% To prove Lemma 1, we need to prove the following\\
% \begin{enumerate}
%     \item The convergence of the objective function, that is $\sum_{i \in \mathcal{N}}f_i(x_i(k)) \rightarrow \sum_{i \in \mathcal{N}}f_i(x_i^*)$.\\
%     \item Satisfying the constraints \eqref{ADMMConstraint} and \eqref{eq:consensus}.
% \end{enumerate}
Lets begin with defining the following error terms: for any $i \in \mathcal{N}$, for $j \in \mathcal{N}_i$, and for $k\in\aleph$,
\begin{eqnarray}
r_{ij}^{k+1} &\triangleq& [x_i^{k+1}-z_{ij}^{k+1}], \label{eq:r}\\
q_{i}^{k+1} &\triangleq& \textstyle \sum_{j \in \mathcal{N}_i}A_{ij}z_{ji}^{k+1}.
\label{eq:q}
\end{eqnarray}

\noindent Under assumption 1, problem \eqref{ADMMFormulation} has at least one optimal solution, denoted by $(x_i^*,z_{ij}^*,\mu_i^*,\lambda_{ij}^*)$ for $i \in \mathcal{N}$ and $j \in \mathcal{N}_i$. Since it satisfies the KKT conditions \cite{boyd2004convex}, we have
\begin{equation*}
    L_{0}(x^*,z^*,\lambda^*,\mu^*) \leq L_{0}(x^{k+1},z^{k+1},\lambda^*,\mu^*),
\end{equation*}
or equivalently,
\begin{align*}
& \sum_{i \in \mathcal{N}}\Big[f_{i}(x_i^*) +\sum_{j \in \mathcal{N}_i}(\lambda_{ij}^*)^{T}(x_i^*-z_{ij}^*) + (\mu_{i}^*)^{T}\sum_{j \in \mathcal{N}_i}A_{ij}z_{ji}^*\Big] \\
  \leq & \sum_{i \in \mathcal{N}}\Big[f_{i}(x_i^{k+1}) + \sum_{j \in \mathcal{N}_i} (\lambda_{ij}^*)^{T}r_{ij}^{k+1} +(\mu_{i}^*)^{T}q_{i}^{k+1} \Big].
\end{align*}
Since the optimal solution satisfies the constraints, we know
$x_{i}^*-z_{ij}^*=0$ and $\sum_{j \in \mathcal{N}_i}  A_{ij}z_{ji}^* = 0$. Hence, the above inequality becomes
\begin{align}
p^* \leq p^{k+1} +  \sum_{i \in \mathcal{N}}  \Big[\sum_{j \in \mathcal{N}_i}(\lambda_{ij}^*)^{T}r_{ij}^{k+1} + (\mu_{i}^*)^{T}q_{i}^{k+1} \Big],
\label{eq:ineq1}
\end{align}
where $p^* =   \sum_{i \in \mathcal{N}} f_{i}(x_i^*)$ and $p^{k+1}=\sum_{i \in \mathcal{N}} f_{i}(x_i^{k+1})$. Also, it follows that
\begin{eqnarray}
\lambda_{ij}^{k+1} - \lambda_{ij}^k & = & (\lambda_{ij}^{k+1} - \lambda_{ij}^*) - (\lambda_{ij}^k - \lambda_{ij}^*), \label{eq1} \\
z_{ij}^{k+1} - z_{ij}^* & = & (z_{ij}^{k+1} - z_{ij}^k) + (z_{ij}^k - z_{ij}^*),  \label{eq2} \\
z_{ij}^{k+1} - z_{ij}^k & = &  (z_{ij}^{k+1} - z_{ij}^*) - (z_{ij}^k - z_{ij}^*),  \label{eq3}\\
\mu_{ij}^{k+1} - \mu_{ij}^k & = & (\mu_{ij}^{k+1} - \mu_{ij}^*) - (\mu_{ij}^k - \mu_{ij}^*). \label{eq4}
\end{eqnarray}

 It follows from \eqref{x-problem} that, for agent $i$,
\begin{align*}
0 =  \nabla_{x_i}f_i(x_i^{k+1}) + \sum_{j \in \mathcal{N}_i}\lambda_{ij}^k + \sum_{j \in \mathcal{N}_i}d_{ij}^k\Big[x_i^{k+1} - z_{ij}^k\Big].
\end{align*}
Substituting \eqref{lambda-problem} into the above equation yields
\begin{align*}
0 = \nabla_{x_i}f_i(x_i^{k+1}) + \sum_{j \in \mathcal{N}_i} \lambda_{ij}^{k+1} + \sum_{j \in \mathcal{N}_i}d_{ij}^k\Big[z_{ij}^{k+1} - z_{ij}^k\Big].
\end{align*}
The above equation implies that $x_i^{k+1}$ also minimizes
\begin{align}
     f_i(x_i^{k+1}) + \sum_{j \in \mathcal{N}_i} \Big[\lambda_{ij}^{k+1} + d_{ij}^k(z_{ij}^{k+1} - z_{ij}^k)\Big]^Tx_i^{k+1}. \label{Alternative1}
\end{align}
\noindent Similarly, it follows from \eqref{z-problem}  that
\begin{align*}
    0 =&  -\lambda_{ij}^{k} - d_{ij}^k \Big[x_i^{k+1} - z_{ij}^{k+1}\Big] \nonumber \\
    & + A_{ji}^T\mu_j^k + w_{j}A_{ji}^T\sum_{\phi \in \mathcal{N}_i}A_{\phi i}z_{i\phi}^{k+1}.
\end{align*}
\noindent Substituting \eqref{mu-problem} and \eqref{lambda-problem} in the above equation yields
\begin{align}
    0 =  -\lambda_{ij}^{k+1} + A_{ji}^T\mu_j^{k+1}.
    \label{Alternative2}
\end{align}
Applying Lagrange duality to \eqref{Alternative1}, we have
\begin{align*}
    & \sum_{i \in \mathcal{N}} \Bigg[ f_i(x_i^{k+1}) + \sum_{j \in \mathcal{N}_i}[\lambda_{ij}^{k+1} + d_{ij}^k (z_{ij}^{k+1} - z_{ij}^{k})]^Tx_i^{k+1} \Bigg]\\
    \leq &  \sum_{i \in \mathcal{N}} \Bigg[ f_i(x_i^*) +  \sum_{j \in \mathcal{N}_i}[\lambda_{ij}^{k+1} + d_{ij}^{k}(z_{ij}^{k+1} - z_{ij}^k)]^Tx_i^*\Bigg].
\end{align*}
It follows from \eqref{Alternative2} that
\begin{align*}
    & \sum_{i \in \mathcal{N}} \sum_{j \in \mathcal{N}_i}\Bigg[-\lambda_{ij}^{k+1} + A_{ji}^T \mu_j^{k+1} \Bigg]^Tz_{ij}^{k+1} \\
    = & \sum_{i \in \mathcal{N}} \sum_{j \in \mathcal{N}_i} \Bigg[-\lambda_{ij}^{k+1} + A_{ji}^T \mu_j^{k+1}\Bigg]^Tz_{ij}^*.
\end{align*}
\noindent Adding the above two expressions together and performing simple manipulations, we obtain
\begin{align}
p^{k+1} - p^* \leq&  \sum_{i \in \mathcal{N}} \sum_{j \in \mathcal{N}_i} \Big[-\lambda_{ij}^{k+1} r_{ij}^{k+1} - (\mu_{j}^{k+1})^TA_{ji}z_{ij}^{k+1} \nonumber \\
& - d_{ij}^k(z_{ij}^{k+1} - z_{ij}^k)^T (r_{ij}^{k+1} + z_{ij}^{k+1} - z_{ij}^*) \Big].
\label{eq:ineq2}
\end{align}
Combining \eqref{eq:ineq1} and \eqref{eq:ineq2} together with \eqref{eq:q} and multiplying both sides by $2$ yield
\begin{align*}
0 \geq & \sum_{i \in \mathcal{N}} \Bigg\{ \sum_{j \in \mathcal{N}_i} \Bigg[ 2(\lambda_{ij}^{k+1} - \lambda_{ij}^*)^T \underbrace{r_{ij}^{k+1}}_{\eqref{lambda-problem}}+ 2d_{ij}^k(z_{ij}^{k+1} - z_{ij}^k)^T \\
& r_{ij}^{k+1} + 2d_{ij}^k (z_{ij}^{k+1} - z_{ij}^k)^T\underbrace{(z_{ij}^{k+1} - z_{ij}^*)}_{\eqref{eq2}} \Big]\\
&+ 2(\mu_i^{k+1}-\mu_i^*)^T(\underbrace{A_{ij}z_{ji}^{k+1})}_{\eqref{mu-problem}}\Bigg]\Bigg\}.
\end{align*}
Performing the substitutions indicated by the underbraces above, we have
\begin{align*}
0 \geq & \sum_{i \in \mathcal{N}} \Bigg[\sum_{j \in \mathcal{N}_i}\Big[ \frac{2}{d_{ij}^k}(\lambda_{ij}^k - \lambda_{ij}^*)^T(\underbrace{\lambda_{ij}^{k+1} - \lambda_{ij}^k)}_{\eqref{eq1}} + d_{ij}^k||\underbrace{r_{ij}^{k+1}}_{\eqref{lambda-problem}}||^2 \\
& + d_{ij}^k||r_{ij}^{k+1} - (z_{ij}^{k+1} - z_{ij}^k)||^2 +d_{ij}^k||\underbrace{z_{ij}^{k+1}-z_{ij}^k}_{\eqref{eq3}}||^2 \\
& + 2d_{ij}^k \underbrace{(z_{ij}^{k+1} - z_{ij}^k)^T}_{\eqref{eq3}}(z_{ij}^k-z_{ij}^*)\Big] + \frac{2}{w_i} (\mu_i^{k+1} - \mu_i^*)^T\\
& (\underbrace{\mu_i^{k+1} - \mu_i^k)}_{\eqref{eq4}}\Bigg].
\end{align*}
Using the substitutions indicated by the underbraces above, we obtain
\begin{align}
0\geq &  \sum_{i \in \mathcal{N}} \Bigg\{\sum_{j \in \mathcal{N}_i} \Bigg[\frac{1}{d_{ij}^k} \Big[||\lambda_{ij}^{k+1} -\lambda_{ij}^*||^2 - ||\lambda_{ij}^k -\lambda_{ij}^*||^2\Big] \nonumber \\
& +  \frac{1}{w_i} \Big[||\mu_i^{k+1}-\mu_i^*||^2 - ||\mu_{i}^k-\mu_i^*||^2\Big]\nonumber \\
&  +  d_{ij}^k\Big[||z_{ij}^{k+1}- z_{ij}^*||^2   - ||z_{ij}^k - z_{ij}^*||^2\Big] +\frac{1}{w_i} ||\mu_{i}^{k+1} \nonumber \\
&- \mu_i^k||^2  + d_{ij}^k\Big[||r_{ij}^{k+1} + (z_{ij}^{k+1} - z_{ij}^k)||^2 \Big] \Bigg]  \Bigg\}. \label{Delta_V}
\end{align}
Considering the following Lyapunov function
\begin{align}
    V(k) = & \sum_{i \in \mathcal{N}} \Bigg\{\sum_{j \in \mathcal{N}_i} \Bigg[d_{ij}^{k} ||z_{ij}^k - z_{ij}^*||^2 + \frac{1}{d_{ij}^k}||\lambda_{ij}^k - \lambda_{ij}^*||^2 \Bigg] \nonumber \\
    & + \frac{1}{w_i}||\mu_i^k - \mu_i^*||^2  \Big]\Bigg\},
\end{align}
we can rewrite inequality \eqref{Delta_V} in terms of the Lyapunov function as
\begin{align}
 V^{k+1} - V^k \leq &  -\sum_{i \in \mathcal{N}} \Bigg[ \sum_{j \in \mathcal{N}_i} d_{ij}^k || r_{ij}^{k+1} + (z_{ij}^{k+1} - z_{ij}^k)||^2 \nonumber\\
 &+ \frac{1}{w_i}||\mu_i^{k+1} - \mu_i^k||^2\Bigg].
  \label{Delta_V2}
\end{align}
Inequality \eqref{Delta_V2} shows that consensus \eqref{eq:consensus} is ensured and that $\mu_i$ converges. Convergence of $\mu_i$ ensures constraint \eqref{ADMMConstraint} is also satisfied. It also follows from \eqref{lambda-problem} that consensus \eqref{eq:consensus} implies convergence of $\lambda_{ij}$. These conclude the proof. {\hfill $\blacksquare$}

%%%%%%%%%%%%%%%%%%%%%%%%%%%%%%%%%%%%%%%%%%%%%%%%%%%%%%%%%%%%%%%%%%%%%%%%%%%%%%%%%%%%%%%%%%
\subsection{Proof of Theorem 1}
It follows from \eqref{dynamic} that, for agent $i$, adaptive gains $d_{ij}$ only appear in \eqref{x-update} but not \eqref{z-update} or \eqref{mu-update} or \eqref{lambda-update}. Hence, the impact of $d_{ij}^k$ on Lyapunov function \eqref{soloLyapunov} can be investigated using its expansion:
\begin{align}
E_i^{k+1}  =& ||\tilde{x}_i^k||^2 - 2\alpha_i(\tilde{x}_i^k)^T \Big[ \nabla_{\tilde{x}_i} f_i(\tilde{x}_i^k)+ \sum_{j \in \mathcal{N}_i} \tilde{\lambda}_{ij}^k \Big]  \nonumber \\
& \boxed{- 2\alpha_i(\tilde{x}_i^k)^T\sum_{j \in \mathcal{N}_i}d_{ij}^k(\tilde{x}_i^k  - \tilde{z}_{ij}^k)} + ||\tilde{\mu}_i^{k+1}||^2 \nonumber \\
& + \sum_{j \in \mathcal{N}_i} \Big[||\tilde{z}_{ji}^{k+1}||^2 + ||\tilde{\lambda}_{ji}^{k+1}||^2 \Big],
\label{soloLyapunov-delta}
\end{align}
in which $\alpha_i^2$ terms are neglected due to $0<\alpha_i<<1$. In \eqref{soloLyapunov-delta}, the boxed sum contains all the terms associated with $d_{ij}^k$. Hence, $E_i^{k+1}$ can assume two different values: one with $d_{ij}^k$ updated according to gain adaptation law  \eqref{ADMM-gains}, and another with no adaptation (i.e., $d_{ij}^k=d_{ij}^{k-1}$ for all $j\in \mathcal{N}_i$). The difference is defined and can be calculated as
\begin{align*}
\Delta E_i^{k+1} := & \; E_i^{k+1}\left|_{\mbox{\scriptsize $d_{ij}^{k}$ updated using \eqref{ADMM-gains}}} - E_i^{k+1}\right|_{\mbox{\scriptsize $d_{ij}^{k}$ not updated}} \\
 = & \; -\epsilon_i^k h_i(\tilde{x},\tilde{z}),
\end{align*}
where $h(\cdot)$ is given by \eqref{ADMM-h}. Thus, the proof is completed by noting that $\Delta E_i^{k+1}<0$ under choice $\epsilon_i^k$ of \eqref{ADMM-epsilon}. \hfill $\blacksquare$

\end{document}